\documentclass[twocolumn,pra,aps,superscriptaddress,showpacs,amsmath,amssymb]{revtex4}

\usepackage{graphicx}
\usepackage{dcolumn}
\usepackage{bm}

\newcommand{\Tr}[1]{\mbox{Tr}\,#1}

\newcommand{\ket}[1]{| #1\rangle}
\newcommand{\bra}[1]{\langle #1|}
\newcommand{\proj}[2]{| #1\rangle\langle #2|}
\newcommand{\ipro}[2]{\langle #1|#2 \rangle}
\newcommand{\expec}[1]{\langle #1 \rangle}
\newcommand{\Eop}{\hat{E}}

\newcommand{\Iop}{\hat{I}}

\newcommand{\Kop}{\hat{K}}

\newcommand{\Mop}{\hat{M}}

\newcommand{\Sop}{\hat{S}}

\newcommand{\Uop}{\hat{U}}
\newcommand{\Vop}{\hat{V}}
\newcommand{\Wop}{\hat{W}}
\newcommand{\Xop}{\hat{X}}
\newcommand{\Yop}{\hat{Y}}

\newcommand{\aop}{\hat{a}}

\begin{document}
\title{Implementing general measurements on linear optical and
solid-state qubits}

\author{Yukihiro Ota}
\affiliation{
Advanced Science Institute, RIKEN, Wako-shi, Saitama, 351-0198, Japan}
\affiliation{
CREST(JST), Kawaguchi, Saitama, 332-0012, Japan}

\author{Sahel Ashhab}
\affiliation{
Advanced Science Institute, RIKEN, Wako-shi, Saitama, 351-0198, Japan}
\affiliation{
Physics Department, University of Michigan, 
Ann Arbor, Michigan 48109-1040, USA}

\author{Franco Nori}
\affiliation{
Advanced Science Institute, RIKEN, Wako-shi, Saitama, 351-0198, Japan}
\affiliation{
Physics Department, University of Michigan, 
Ann Arbor, Michigan 48109-1040, USA}

\date{\today}

\begin{abstract}
We show a systematic construction for implementing general measurements
 on a single qubit, including both strong (or projection) and weak
 measurements. 
We mainly focus on linear optical qubits. 
The present approach is composed of simple and feasible elements, i.e.,
 beam splitters, wave plates, and polarizing beam splitters. 
We show how the parameters characterizing the measurement
operators are controlled by the linear optical elements. 
We also propose a method for the implementation of general
 measurements in solid-state qubits. 
\end{abstract}


\maketitle

\section{Introduction}
The description of measurement in quantum
mechanics\,\cite{vonNeumann:1955,Davies:1976,Kraus:1983,Peres:1993} is
formulated as an operation to extract information from a quantum system.  
This operation causes a disturbance to the system. 
This means that the measurement accuracy (or the amount of information
extracted from the system) is closely related to
the
back-action\,\cite{Braginsky;Khalili:1992,Banaszek:2001,Ozawa:2004,Clerk;Schoelkopf:2010,Wiseman;Milburn:2010}.   
This is a curious feature of measurement processes in
quantum mechanics. 
Such a character is actively used in measurement-based methods for
quantum
engineering (e.g., Refs.\,\cite{Nakazato;Yuasa:2003,Koashi;Ueda:1999,Korotkov;Keane:2010,Ashhab;Nori:2010,Paraoanu:2011:EPL,Paraoanu:2011:FP,Ota;Nori}).  

A general (or weak) measurement is associated with a positive-operator
valued measure (POVM)\,\cite{Davies:1976}. 
A POVM on a measured quantum system can be expressed as a
projection-valued measure in an extended system including the target
system and an ancillary (or probe) system, as seen in, e.g.,
Ref.\,\cite{Peres:1993}. 
Therefore, we may construct arbitrary measurements from von Neumann
measurements on the extended system. 
However, this statement does not give a specific and simple recipe
to design measurement operators. 
Hence, a systematic approach to realize this general idea in
specific physical systems is highly desirable. 

A large number of experimental studies on general
measurements have been performed. 
Huttner {\it et al.}\,\cite{Huttner;Gisin:1996} discriminated with two
nonorthogonal states. 
Gillett {\it et al.}\,\cite{Gillett;White:2010} demonstrated quantum
feedback control for a photonic polarization qubit. 
A polarizing beam splitter with a tunable reflection coefficient was
used on the above-mentioned two experiments. 
Kwiat {\it et al.}\,\cite{Kwiat;Gisin:2001} implemented an entanglement
concentration protocol with a partial-collapse measurement in a photonic
qubit. 
Kim {\it et al.}\,\cite{Kim;Kim:2009} demonstrated a reversal operation
of a weak measurement for a photonic qubit. 
The key idea in the two experiments is to use Brewster-angle glass
plates. 
Katz {\it et al.}\,\cite{Katz;Korotkov:2008} performed a conditional
recovery of a quantum state with a partial-collapse measurement in
Josephson phase qubits. 
Iinuma {\it et al.}\,\cite{Iinuma;Hofmann:2011} studied the observation
of a weak value. 
Kocsis {\it et al.}\,\cite{Kocsis;Steinberg:2011} examined a photon's
``trajectories'' through a double-slit interferometer with weak
measurements of the photon momentum. 

In this paper, we show a method to implement general measurements on a
single qubit, including both von Neumann and weak measurements. 
We mainly focus on a linear optical qubit. 
Depending on the path degree of freedom in an interferometer, a
polarization state is transformed by a non-projective positive operator.  
We develop the idea proposed in Ref.\,\cite{Iinuma;Hofmann:2011} and
give a systematic prescription for designing various measurement operators. 
The present approach is composed of simple and basic linear optical
elements, i.e., beam splitters, wave plates, and polarizing beam splitters. 
We show how the parameters characterizing the measurement
operators (e.g., the measurement strength) can be tuned. 
Furthermore, we show a method for implementing general measurements on
a solid-state qubit. 

The paper is organized as follows. 
In Sec.\,\ref{sec:general}, we show the basic idea for constructing
general measurements on linear optical qubits.   
The notation used in this paper is explained, there. 
In Sec.\,\ref{sec:los}, we propose systematic approaches for
implementing measurement operators.  
These method are simple and could be implementable in experiments. 
In addition, in Sec.\,\ref{sec:sss}, we study how a general measurement
can be implemented in solid-state qubits. 
Section \ref{sec:summary} is devoted to a summary of the ideas and the
results. 

\section{Setting}
\label{sec:general}
Let us consider a linear optical system with $N$ input and $N$ output
modes.  
The following arguments are also applicable to other physical systems
such as neutron interferometers\,\cite{Sponar;Hasagawa:2010}.  
The annihilation (creation) operator in the $n$th input mode is defined
as $\aop_{{\rm in},n}$ ($\aop_{{\rm in},n}^{\dagger}$) with
$n=1,\,2,\ldots,N$.  
The vacuum $\ket{0}$ is defined as $\aop_{{\rm in},n}\ket{0}=0$. 
The bosonic canonical commutation relations are 
\(
[\aop_{{\rm in},m},\aop_{{\rm in},n}^{\dagger}]=\delta_{nm}
\) 
and  
\(
[\aop_{{\rm in},m},\aop_{{\rm in},n}]
=
[\aop_{{\rm in},m}^{\dagger},\aop_{{\rm in},n}^{\dagger}]
=0
\). 
Similarly, we define the annihilation (creation) operator in the $n$th
output mode as $\aop_{{\rm out},n}$ ($\aop_{{\rm out},n}^{\dagger}$). 
A single photon with arbitrary polarization enters the system 
from the first input mode. 
The polarization is described by a vector
$\ket{\psi}\in\mathbb{C}^{2}$, with $\ipro{\psi}{\psi}=1$. 
We assume that the input state is a pure state
\(
 \ket{\Psi} 
= \ket{\psi} \otimes \aop_{{\rm in},1}^{\dagger}\ket{0}
\). 
In this setting, the photon polarization is measured, using the path
degree of freedom in the interferometer as an ancillary qubit. 
We note that the arguments in this section can be straightforwardly
extended to the case of a mixed initial state of polarization.  

We calculate the photon state at each output mode. 
The field operators at the output modes are related to the input modes
via a linear canonical
transformation\,\cite{vanHemmen:1980,Umezawa:1993,Loudon:2000,Kok;Milbrun:2007} 
because the system is composed of linear optical elements (e.g., beam
splitters).  
The polarization of the photon at the $n$th output mode is described by 
\(
\ket{\phi_{n}}
=
\bra{0}\aop_{{\rm out},n}\ket{\Psi}
\). 
Thus, we find that in terms of the output mode operators the photon
state is written as 
\(
\ket{\Psi^{({\rm out})}} 
= 
\sum_{n=1}^{N} \ket{\phi_{n}} \otimes 
\aop_{{\rm out},n}^{\dagger}\ket{0}
\). 
Since we assume that there is no photon loss, we have 
\(
\expec{\Psi^{\rm out} |\Psi^{\rm out}}=\expec{\Psi |\Psi}
\). 
The linearity of the system indicates that there exist linear
operators $\Xop_{n}$ on $\mathbb{C}^{2}$ such that 
\(
\ket{\phi_{n}} = \Xop_{n}\ket{\psi}
\). 
The normalization condition $\ipro{\Psi}{\Psi}=1$ implies that 
\(
\sum_{n=1}^{N}\Xop_{n}^{\dagger}\Xop_{n} = \Iop_{2}
\), where the identity operator on $\mathbb{C}^{2}$ is $\Iop_{2}$. 
Therefore, we find that the initial state $\ket{\psi}$ is transformed by
a linear operator which is related with an element of a POVM. 

It is convenient for designing various kinds of POVM to add another
linear optical system to some of the output modes in the interferometer
before the photon detection. 
For example, let us set wave plates in each output mode. 
The wave plates in the $n$th mode describe a unitary gate $V_{n}$ given
as the unitary part of $\Xop_{n}$. 
Using the right-polar decomposition\,\cite{Horn;Johnson:1985}, we
find that 
\(
\Xop_{n} = \Vop_{n}^{\dagger}\Mop_{n}
\), where 
\(
\Mop_{n}=(\Xop_{n}^{\dagger}\Xop_{n})^{1/2}
\) is a positive operator. 
The role of the wave plates is to remove the effect of the unitary
evolution $\Vop_{n}^{\dagger}$ from $\Xop_{n}$. 
Using the application of such ``path-dependent'' unitary operators, we
find that 
\(
\ket{\Psi^{(\rm out)}} 
\to 
\sum_{n=1}^{N} \ket{\psi_{n}} \otimes \aop_{{\rm out},n}^{\dagger}\ket{0}
\), where 
\(
 \ket{\psi_{n}}
=
\Vop_{n}\ket{\phi_{n}}
=
\Mop_{n}\ket{\psi}
\). 
The measurement operators $\Mop_{n}$ satisfy  
\(
 \sum_{n=1}^{N}\Mop_{n}^{\dagger}\Mop_{n}=\Iop_{2}
\). 
Thus, we have the measurements corresponding to the POVM 
\(
\{\Eop_{n}\}_{n=1}^{N}
\) with $\Eop_{n}=\Mop_{n}^{\dagger}\Mop_{n}$. 
The positive operator $\Mop_{n}$ is the essential part (i.e., the
back-action associated with the measurement) of non-unitarity of 
$\Xop_{n}$ and called minimally-disturbing measurement, following
Ref.\,\cite{Wiseman;Milburn:2010}. 
Throughout this paper, we will focus on such minimally-disturbing
measurements. 

Let us characterize the measurement operators $\Mop_{n}$. 
We can expand $\Mop_{n}$ as 
\(
\Mop_{n}
=
\sum_{i=1}^{2} m_{n,i}\proj{m_{n,i}}{m_{n,i}}
\), with the eigenvalues $m_{n,i}\,(\ge 0)$ and the associated
eigenvectors $\ket{m_{n,i}}$. 
The eigenvalues are related to the measurement strength, while the
eigenvectors are regarded as the measurement direction.  
Let us consider the case when $m_{n,1}=1$ and $m_{n,2}=0$, for example. 
We find that $\Mop_{n}$ is a projection operator (i.e., a sharp
measurement) in the direction of $\vec{m}_{n,1}$,  
where $\vec{m}_{n,1}$ is the Bloch vector corresponding to
$\proj{m_{n,1}}{m_{n,1}}$. 
The indistinguishability between the elements of a POVM is
characterized by the Hilbert-Schmidt inner product 
\(
\Tr \Eop_{n}^{\dagger}\Eop_{m}
\) for $n\neq m$. 
The elements of a projection-valued measure are distinguishable (i.e., 
the inner products are zero), for example. 
Now, we pose the question: How are these measurement
features controlled by typical linear optical devices? 
We will answer this question in the next section. 

\begin{figure}[htbp]\centering
\scalebox{0.43}[0.43]{\includegraphics{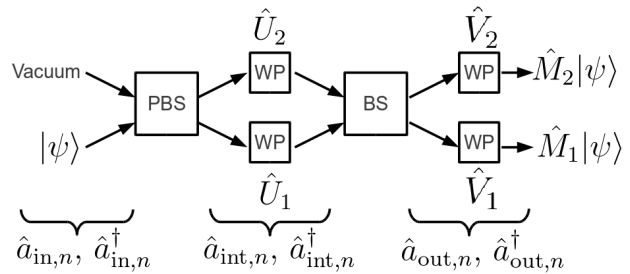}}
\caption{Schematic diagram of a proposal for implementing symmetric,
 arbitrary-strength two-outcome measurements on a single-photon's
 polarization. 
This setup is essentially equal to the one in Ref.\,\cite{Iinuma;Hofmann:2011}. 
Three kinds of field operators (i.e., 
$\aop_{{\rm in},n}$, $\aop_{{\rm int},n}$, and 
$\aop_{{\rm out},n}$) are needed for representing the path degrees of
 freedom. A single photon
 with arbitrary polarization (described by $\ket{\psi}$) enters a
 polarizing beam splitter (PBS). Next, the polarization is 
transformed by a wave plate (WP) in each arm. Then, the two arms are 
 recombined in a beam splitter (BS). After 
performing unitary gates $\Vop_{1}$ and $\Vop_{2}$ for compensation
 purposes, the resultant polarization states are expressed in terms of
 the measurement operators $\Mop_{1}$ and $\Mop_{2}$ given in
 Eqs.\,(\ref{eq:def_elm1}) and (\ref{eq:def_elm2}).}
\label{fig:diagram1} 
\end{figure}
\begin{figure}[htbp]
\centering
\scalebox{0.45}[0.45]{\includegraphics{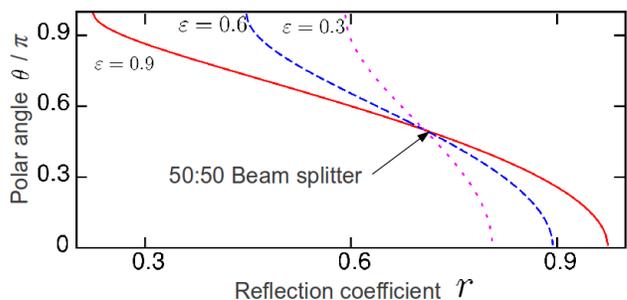}}
\caption{(Color online) Measurement direction of the measurement
 operators (\ref{eq:def_elm1}) and (\ref{eq:def_elm2}) for 
the measurement strength $\varepsilon=0.9$ (red solid), $0.6$
 (blue dashed), and $0.3$ (magenta dotted). 
The horizontal axis represents the reflection coefficient $r$ of the  
 beam splitter, while the vertical axis is the polar angle $\theta$ of
 the measurement direction. All curves merge at 
$(r,\,\theta)=(1/\sqrt{2},\,\pi/2)$. This case corresponds to the 
 use of a $50:50$ beam splitter in Fig.\,\ref{fig:diagram1}.}
\label{fig:rc_theta}
\end{figure}
\begin{figure}[htbp]
\centering
\scalebox{0.46}[0.46]{\includegraphics{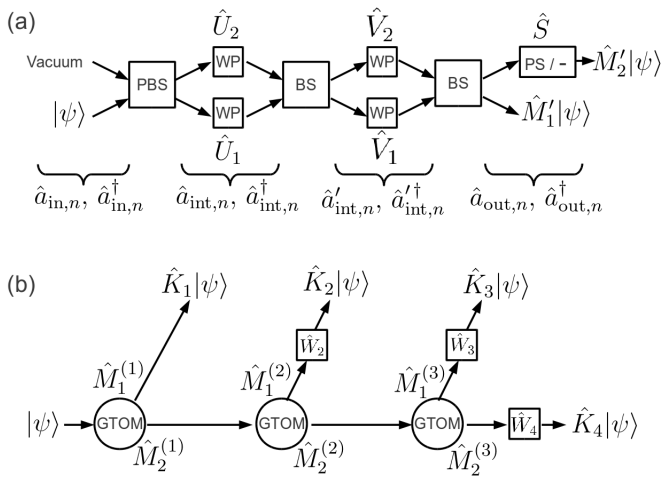}}
\caption{Schematic diagrams of the construction for general measurements
 with (a) two  outcomes and (b) $N$ outcomes ($N=4$). In (a), the
 interferometer for a symmetric arbitrary-strength two-outcome
 measurement followed by a beam splitter leads to general two-outcome
 measurements. The final box leading to $\Mop_{2}^{\prime}\ket{\psi}$
 after the last beam splitter  represents a unitary gate for
 compensation purposes. Depending on system's parameters, this unitary
 gate becomes either a phase-shift gate 
 or the identity operator (i.e., does nothing). In (b), the repeated 
 application of general two-outcome measurements (GTOM) allows the
 construction of a POVM with multiple outcomes.} 
\label{fig:diagram2} 
\end{figure}

\section{Design of general measurements on linear optical qubits}
\label{sec:los}
\subsection{Symmetric arbitrary-strength two-outcome measurements}
\label{subsec:sas}
Let us first define a special class of general measurements, which is
the target operation in this subsection. 
We consider a symmetric two-outcome POVM in $\mathbb{C}^{2}$ composed of
convex combinations of two orthogonal projection operators
\begin{eqnarray}
&&
\Mop_{1}
=
\sqrt{\frac{1 + \varepsilon}{2}}
\proj{m_{+}}{m_{+}}
+
\sqrt{\frac{1 - \varepsilon}{2}}
\proj{m_{-}}{m_{-}} , \label{eq:def_elm1}\\
&&
\Mop_{2}
=
\sqrt{\frac{1 - \varepsilon}{2}}
\proj{m_{+}}{m_{+}}
+
\sqrt{\frac{1 + \varepsilon}{2}}
\proj{m_{-}}{m_{-}} ,
\label{eq:def_elm2} 
\end{eqnarray}
with $0 \le \varepsilon \le 1$ and 
\(
\ipro{m_{+}}{m_{-}}=0
\). 
Adjusting a real parameter $\varepsilon$ allows arbitrary measurement
strength. 
We find that if the coefficient for 
$\proj{m_{+}}{m_{+}}$ increases, then the one for
$\proj{m_{-}}{m_{-}}$ decreases, and vice versa. 
Thus, these measurement operators are symmetric or balanced with respect
to the parameter $\varepsilon$. 
We call the measurements described by these linear operators
symmetric arbitrary-strength two-outcome measurments (SASTOM) on
$\mathbb{C}^{2}$. 
The disturbance induced by a SASTOM is considered
to be minimum, as shown in Ref.\,\cite{Banaszek:2001}. 
Another important property of a SASTOM is 
\(
\Tr \Mop_{1}^{\dagger}\Mop_{1} 
= 
\Tr \Mop_{2}^{\dagger}\Mop_{2}
=
1
\). 
An application of a SASTOM to quantum protocols is shown in, e.g.,
Ref.\,\cite{Ota;Nori}. 

Now, we propose a systematic way to construct this measurement. 
Let us consider an interferometer with two input and two output
modes, as shown in Fig.\,\ref{fig:diagram1}.  
A single photon enters a polarizing beam
splitter\,\cite{Meschede:2007} from the first input mode. 
Subsequently, its polarization at each arm is transformed by wave
plates.  
Then, the two modes are recombined in a beam splitter. 
This system is essentially the same as in
Ref.\,\cite{Iinuma;Hofmann:2011}. 
We reformulate the construction manner of a SASTOM
in Ref.\,\cite{Iinuma;Hofmann:2011} and extend it to make
more general types of measurements. 
Unlike the proposal in Ref.\,\cite{Iinuma;Hofmann:2011}, our proposals
use beam splitters with tunable reflection coefficients. 

Let us apply the basic idea in Sec.\,\ref{sec:general} to this
interferometer. 
We need three kinds of field operators for describing the path degrees of
freedom, i.e., input, intermediate, and output field operators, as shown in
Fig.\,\ref{fig:diagram1}. 
We have an initial pure state 
\begin{equation*}
 \ket{\Psi^{({\rm in})}}
=
\ket{\psi} \otimes \aop_{{\rm in},1}^{\dagger}\ket{0},
\quad
\ket{\psi}
=c_{\rm H}\ket{{\rm H}} + c_{\rm V}\ket{{\rm V}}, 
\end{equation*}
with the horizontal polarization state $\ket{\rm H}$ and the vertical
polarization state $\ket{\rm V}$. 
The field operators $\aop_{{\rm int},n}$ and 
$\aop_{{\rm int},n}^{\dagger}$ at the output modes of the polarizing
beam splitter are related to $\aop_{{\rm in},n}$ and
$\aop_{{\rm in},n}^{\dagger}$ via 
\(
\aop_{{\rm int,H},1} 
= 
\aop_{{\rm in,H},1}
\), 
\(
\aop_{{\rm int,H},2} 
= 
\aop_{{\rm in,H},2}
\), 
\(
\aop_{{\rm int,V},1} 
= 
\aop_{{\rm in,V},2}
\), 
and 
\(
\aop_{{\rm int,V},2} 
= 
\aop_{{\rm in,V},1}
\)\,\cite{Kok;Milbrun:2007}, 
where 
\(
\aop_{{\rm in,H},1} = \proj{{\rm H}}{{\rm H}}\otimes 
\aop_{{\rm in},1}
\), etc. 
After the single photon passes through the polarizing beam splitter, we
find that the photon state is written as 
\(
 \ket{\Psi^{({\rm int})}}
=
\ket{u_{1}}\otimes \aop_{{\rm int,}1}^{\dagger}\ket{0}
+
\ket{u_{2}}\otimes \aop_{{\rm int,}2}^{\dagger}\ket{0}
\) 
with 
\(
\ket{u_{1}}
=
c_{\rm H}\ket{{\rm H}}
\) and 
\(
\ket{u_{2}}
=
c_{\rm V}\ket{{\rm V}}
\). 
We remark that the polarizing beam splitter produces entanglement
between the polarization and the path degrees of freedom in the
interferometer.  
Next, we perform unitary operations for the polarization on
each of these intermediate modes with the wave plates.  
The use of half- and quarter-wave plates allows the construction of
arbitrary elements of
$\mbox{SU}(2)$\,
\cite{Simon;Mukunda:1990,Bhadari;Dasgupta:1990}. 
Hence, using the path-dependent wave plates, we find that 
\begin{equation*}
 \ket{\Psi^{(\rm int)}} 
\to  
\sum_{n=1}^{2}\Uop_{n}\ket{u_{n}}
\otimes \aop_{{\rm int,}n}^{\dagger}\ket{0}, 
\end{equation*}
with $\Uop_{n}\in \mbox{SU}(2)$. 
The canonical transformation associated with the beam
splitter\,\cite{Loudon:2000,Kok;Milbrun:2007} 
is 
\(
 \aop_{{\rm out},1} 
= r \aop_{{\rm int,}1} + t \aop_{{\rm int,}2}
\)
and 
\(
 \aop_{{\rm out},2} 
= t \aop_{{\rm int,}1} - r \aop_{{\rm int,}2}
\), 
where $r^{2}+t^{2}=1$ and $0\le r,t \le 1$. 
After the single photon passes through the beam splitter, we find that
the photon state is expressed by 
\begin{equation*}
 \ket{\Psi^{(\rm out)}}
=
\sum_{n=1}^{2}
\ket{\phi_{n}} \otimes \aop_{{\rm out},n}^{\dagger}\ket{0}, 
\end{equation*}
where 
\(
 \ket{\phi_{1}}
=
r \Uop_{1}\ket{u_{1}}
+
t \Uop_{2}\ket{u_{2}}
\) 
and 
\(
 \ket{\phi_{2}}
=
 t \Uop_{1}\ket{u_{1}}
-
 r \Uop_{2}\ket{u_{2}}
\). 
We find that the linear operators satisfying 
\(
\ket{\phi_{n}} = \Xop_{n}\ket{\psi}
\) are 
\(
\Xop_{1}
= 
r\Uop_{1}\proj{{\rm H}}{{\rm H}} 
+ t \Uop_{2}\proj{{\rm V}}{{\rm V}}
\) and 
\(
\Xop_{2}
=
t \Uop_{1}\proj{{\rm H}}{{\rm H}}
- r \Uop_{2}\proj{{\rm V}}{{\rm V}}
\). 
The positive operators associated with $\Xop_{1}$ and $\Xop_{2}$ are
given, respectively, by Eqs.\,(\ref{eq:def_elm1}) and
(\ref{eq:def_elm2}) with  
\begin{equation}
 \varepsilon
=
\sqrt{
1
-
4r^{2}t^{2}
(1-|w|^{2})
},
\quad
w 
= \expec{ {\rm H} | \Uop_{1}^{\dagger}\Uop_{2} | {\rm V}}. 
\label{eq:def_eps}
\end{equation}
The definition of $w$ implies that the choice of the unitary gates
$\Uop_{1}$ and $\Uop_{2}$ is not unique for constructing the measurement
operators $\Mop_{n}$. 
For example, one can set $\Uop_{1}$ as the identity operator and still
obtain any desired value of $w$ by adjusting $\Uop_{2}$. 
Iinuma {\it et al.}\,\cite{Iinuma;Hofmann:2011} set
$\Uop_{1}=\Uop_{2}^{\dagger}$, which results in a real value for $w$ and
thus imposes a constraint on the measurement operators that can be
constructed. 
The range of $|w|$ is 
\(
0 \le |w| \le 1
\) since $w$ is an off-diagonal element of a unitary operator.  
The basis vectors $\ket{m_{+}}$ and $\ket{m_{-}}$ for $w \neq 0$ are 
\begin{eqnarray}
&&
\ket{m_{+}}
=
\cos\frac{\theta}{2}\ket{{\rm H}} 
+ 
e^{-i\phi/2}
\sin\frac{\theta}{2}\ket{{\rm V}}, 
\label{eq:def_mplus}\\
&&
\ket{m_{-}}
=
-e^{i\phi/2}\sin\frac{\theta}{2}\ket{{\rm H}}
+\cos\frac{\theta}{2} \ket{{\rm V}}, 
\label{eq:def_mminus}
\end{eqnarray}
where 
\(
\tan(\theta/2)
=
(t^{2} - r^{2} + \varepsilon)/2rt |w|
\)
and 
\(
e^{i\phi/2}=w/|w|
\). 
When $w=0$ and $r^{2} \ge t^{2}$ ($r^{2} < t^{2}$), we have 
\(
\ket{m_{+}} = \ket{{\rm H}}
\) and 
\(
\ket{m_{-}} = \ket{{\rm V}}
\) 
(\(
\ket{m_{+}} = \ket{{\rm V}}
\) and 
\(
\ket{m_{-}} = -\ket{{\rm H}}
\)). 
The fact that \(
[\Xop_{1}^{\dagger}\Xop_{1},
\Xop_{2}^{\dagger}\Xop_{2}]=0
\) indicates that $\Mop_{1}$ and $\Mop_{2}$ are simultaneously
diagonalizable. 
The expression for $\Vop_{n}$ such that
$\Xop_{n}=\Vop_{n}^{\dagger}\Mop_{n}$ is calculated straightforwardly. 
After the applications of $\Vop_{n}$, we find that 
\begin{equation*}
 \ket{\Psi^{({\rm out})}}
\to 
\sum_{n=1}^{2}
\Mop_{n}\ket{\psi} \otimes \aop_{{\rm out},n}^{\dagger}\ket{0}. 
\end{equation*}
The tunable parameter $\varepsilon$ is the measurement
strength and completely determines the
indistinguishability between the POVM elements, 
\(
\Tr \Eop_{1}^{\dagger}\Eop_{2}
= (1-\epsilon^{2})/2
\). 
Thus, we have shown that the interferometer drawn in
Fig.\,\ref{fig:diagram1} is a measurement apparatus for performing
a SASTOM.  

Let us show how the SASTOM can be adjusted via the
linear optical elements. 
The interferometer contains three independent control parameters: the
reflection coefficient $r$, the modulus of $w$, and the phase of $w$.  
The latter two parameters are related to the wave plates. 
The measurement direction is characterized by the Bloch vector 
\(
(
\sin\theta \cos\phi, \sin\theta \sin \phi, \cos\theta
)
\). 
For the calculation of the Bloch vector the Pauli matrices
are defined as 
\(
\hat{\sigma}_{x} = \proj{{\rm H}}{{\rm V}} + \proj{{\rm V}}{{\rm H}}
\), 
\(
\hat{\sigma}_{y} = -i\proj{{\rm H}}{{\rm V}} + i\proj{{\rm V}}{{\rm H}}
\), and 
\(
\hat{\sigma}_{z} = \proj{{\rm H}}{{\rm H}} - \proj{{\rm V}}{{\rm V}}
\). 
Since the azimuthal angle $\phi$ is equal to the phase of $w$, this
quantity is controlled at will via phase-shift gates. 
The polar angle $\theta$ and the measurement strength $\varepsilon$ are
functions of $r$ and $|w|$. 
We can find that $0 \le \theta \le \pi$ and $0 \le \varepsilon \le 1$
when changing $r$ and $|w|$ independently. 
We evaluate $\theta$ as a function of $r$ and $\varepsilon$. 
Figure \ref{fig:rc_theta} shows that we can take an arbitrary
measurement direction for given $\varepsilon$. 

The comparison to the method in Ref.\,\cite{Iinuma;Hofmann:2011} is
useful for understanding our proposal for a SASTOM. 
Let us consider the case when the beam splitter is a $50:50$ beam
splitter ($r=1/2$) and the unitary operators for the wave plates are 
$\Uop_{1}=\exp(-i2\eta \hat{\sigma}_{y})$ 
and  
$\Uop_{2} = \exp(i2\eta \hat{\sigma}_{y})$. 
We find that $\varepsilon = |w|=|\sin(4\eta)|$, 
\(
\theta = \frac{\pi}{2}
\), and $\phi=0$. 
The measurement direction is fixed, and it is characterized by 
\(
\ket{m_{+}} = (\ket{{\rm H}} + \ket{{\rm V}})/\sqrt{2}
\)
and 
\(
\ket{m_{-}} = (-\ket{{\rm H}} + \ket{{\rm V}})/\sqrt{2}
\). 
In fact, all curves in Fig.\,\ref{fig:rc_theta} merge at 
\(
(r,\,\theta)=(1/\sqrt{2},\,\pi/2)
\). 
Thus, the only tunable parameter in Ref.\,\cite{Iinuma;Hofmann:2011} is
$|w|$. 
Two additional real parameters are necessary for controlling
the measurement direction of a SASTOM on a single qubit. 
For this purpose, our proposal uses the tunable reflection coefficient
in the beam splitter and the phase of $w$. 
Alternatively, the measurement direction can be controlled by using
additional wave plates (i.e., a unitary operator) before the polarizing
beam splitter. 

\subsection{General two-outcome measurements}
Various quantum protocols with measurement operators not expressed
by Eqs.\,(\ref{eq:def_elm1}) and (\ref{eq:def_elm2}) have been proposed
(e.g., Refs.\,\cite{Korotkov;Keane:2010,Ashhab;Nori:2010}). 
In the remaining parts of this section, we extend the approach 
developed in Sec.\,\ref{subsec:sas} for implementing such general
measurements. 
Two generalization routes may exist. 
One involves increasing the number of parameters characterizing the
two measurement operators $\Mop_{1}$ and $\Mop_{2}$. 
The other is to increase the number of outcomes. 
First, we examine the former. 
The eigenvalues of the measurement operators (\ref{eq:def_elm1}) and
(\ref{eq:def_elm2}) are parametrized by $\varepsilon$. 
The measurement direction contains the two parameters $\theta$ and
$\phi$. 
Thus, the number of parameters in the measurement operators
is equal to a density matrix on $\mathbb{C}^{2}$. 
This point is also confirmed by the fact that  
\(
\Tr \Mop_{n}^{\dagger}\Mop_{n} = 1
\). 
Since the positive operator $\Mop_{n}^{\dagger}\Mop_{n}$ is a density
matrix on $\mathbb{C}^{2}$, its square root $\Mop_{n}$ is characterized
by three real parameters. 

We consider the interferometer shown in Fig.\,\ref{fig:diagram2}(a). 
The main difference with Fig.\,\ref{fig:diagram1} is that the present
system has an additional beam splitter with reflection coefficient
$r^{\prime}$ and transmission coefficient $t^{\prime}$. 
Accordingly, we need four kinds of field operators for describing the
path degrees of freedom. 
The calculations before the last beam splitter are the same as in
Sec.\,\ref{subsec:sas}. 
Namely, the first polarizing beam splitter creates entanglement between
the polarization and the path degree of freedom. 
The subsequent wave plates transform photon's polarization through
unitary operators depending on the path degree of freedom. 
The corresponding unitary operator $\Uop_{n}$ is an arbitrary element of
$\mbox{SU}(2)$, as seen in Sec.\,\ref{subsec:sas}. 
Then, the two paths are recombined in the intermediate beam splitter
with reflection coefficient $r$. 
At this point, the polarization state in the $n$th mode is described by
$\Xop_{n}\ket{\psi}$. 
After the photon passes through this intermediate beam splitter, the
unitary operators $\Vop_{1}$ and $\Vop_{2}$ are applied to the first and
the second paths, respectively, in order to remove the unitary parts of
$\Xop_{1}$ and $\Xop_{2}$. 
These unitary operators $\Vop_{1}$ and $\Vop_{2}$ are automatically
determined when calculating the right-polar decompositions of $\Xop_{1}$
and $\Xop_{2}$.  
Thus, the resultant state in the $n$th mode becomes
$\Mop_{n}\ket{\psi}$, as seen in Eqs.\,(\ref{eq:def_elm1}) and
(\ref{eq:def_elm2}). 
As shown in Sec.\,\ref{subsec:sas}, in this step, we have three tunable
parameters, i.e., $r$, the modulus of $w$, and the phase of $w$, where 
\(
w = \expec{{\rm H}|\Uop_{1}^{\dagger}\Uop_{2}|{\rm V}}
\). 

Let us now consider what happens in the additional part. 
We write the input (output) field operators in the last beam splitter as 
\(
\aop_{{\rm int},n}^{\prime}
\) 
and 
\(
\aop_{{\rm int},n}^{\prime\,\dagger}
\) 
(\(
\aop_{{\rm out},n}
\) 
and 
\(
\aop_{{\rm out},n}^{\dagger}
\)). 
The related linear canonical transformations is  
\(
 \aop_{{\rm out},1}
=
r^{\prime}\aop_{{\rm int},1}^{\prime}
+
t^{\prime}\aop_{{\rm int},2}^{\prime}
\) and 
\(
 \aop_{{\rm out},2}
=
t^{\prime}\aop_{{\rm int},1}^{\prime}
-
r^{\prime}\aop_{{\rm int},2}^{\prime}
\) 
with $(r^{\prime})^{2} + (t^{\prime})^{2}=1$ and 
$0\le r^{\prime},t^{\prime} \le 1$. 
After the photon passes through the last beam splitter, its state
is written as 
\(
\ket{\Psi^{({\rm out})}}
=
\sum_{n=1}^{2} 
\Xop_{n}^{\prime}\ket{\psi}
\otimes \aop_{{\rm out},n}^{\dagger}\ket{0}
\), 
where 
\(
\Xop_{1}^{\prime}
=
r^{\prime}\Mop_{1} + t^{\prime}\Mop_{2}
\) 
and 
\(
\Xop_{2}^{\prime}
=
t^{\prime}\Mop_{1} - r^{\prime}\Mop_{2}
\). 
In other words, we find that 
\begin{widetext}
\begin{eqnarray*}
&&
\Xop_{1}^{\prime}
=
\left(
r^{\prime}\sqrt{\frac{1+\varepsilon}{2}}
+
t^{\prime}\sqrt{\frac{1-\varepsilon}{2}}
\right) 
\proj{m_{+}}{m_{+}} 
+
\left(
r^{\prime}\sqrt{\frac{1-\varepsilon}{2}}
+
t^{\prime}\sqrt{\frac{1+\varepsilon}{2}}
\right) 
\proj{m_{-}}{m_{-}}, \\
&&
\Xop_{2}^{\prime}
=
\left(
t^{\prime}\sqrt{\frac{1+\varepsilon}{2}}
-
r^{\prime}\sqrt{\frac{1-\varepsilon}{2}}
\right) 
\proj{m_{+}}{m_{+}} 
+
\left(
t^{\prime}\sqrt{\frac{1-\varepsilon}{2}}
-
r^{\prime}\sqrt{\frac{1+\varepsilon}{2}}
\right) 
\proj{m_{-}}{m_{-}}, 
\end{eqnarray*}
\end{widetext}
where $\varepsilon$, $\ket{m_{+}}$, and $\ket{m_{-}}$ are given in
Eqs.\,(\ref{eq:def_eps}), (\ref{eq:def_mplus}), and
(\ref{eq:def_mminus}), respectively. 
The positive-operator parts of $\Xop_{1}^{\prime}$ and
$\Xop_{2}^{\prime}$ are, respectively, 
\begin{eqnarray}
&&
\Mop_{1}^{\prime} 
=
\sqrt{p}
\proj{m_{+}}{m_{+}}
+
\sqrt{q}
\proj{m_{-}}{m_{-}}, \label{eq:def_gwm1}\\
&&
\Mop_{2}^{\prime} 
=
\sqrt{1-p}
\proj{m_{+}}{m_{+}}
+
\sqrt{1-q}
\proj{m_{-}}{m_{-}}.\label{eq:def_gwm2}
\end{eqnarray}
The measurement direction is the same as the SASTOM in the
previous subsection. 
This means that the basis vectors $\ket{m_{+}}$ and $\ket{m_{-}}$ do not
depend on the parameter $r^{\prime}$. 
We remark that $\Mop_{1}^{\prime}=\Xop_{1}^{\prime}$ since
$\Xop_{1}^{\prime}$ is a positive operator. 
The measurement operator $\Mop_{2}^{\prime}$ is related with the linear
operator $\Xop_{2}^{\prime}$ via a unitary operator, 
\(
\Mop_{2}^{\prime} = \Sop \Xop_{2}^{\prime}
\), with $\Sop^{\dagger}\Sop=\Sop\Sop^{\dagger}=\Iop_{2}$. 
With some algebra one can show that when 
\begin{equation*}
 \sqrt{1-\varepsilon^{2}} \le 2r^{\prime}t^{\prime},
\end{equation*}
the unitary operator $\Sop$ is a phase-shift gate (i.e., 
$\Sop = \proj{m_{+}}{m_{+}} - \proj{m_{-}}{m_{-}}$). 
Otherwise, $\Sop$ is equal to the identity operator, up to an
overall phase. 
The measurement operators $\Mop_{1}^{\prime}$ and $\Mop_{2}^{\prime}$
are characterized by two independent positive parameters $p$ and $q$ 
($0 \le p,q\le 1$). 
In contrast to a SASTOM, the trace of
$\Mop_{n}^{\prime\,\dagger}\Mop_{n}^{\prime}$ is not fixed. 
We find that 
\(
\Tr \Mop_{1}^{\prime\,\dagger}\Mop_{1}^{\prime} 
=
1 + \Delta
\) 
and 
\(
\Tr \Mop_{2}^{\prime\,\dagger}\Mop_{2}^{\prime} 
=
1 - \Delta
\), with 
\(
\Delta = p+q-1
\). 
The indistinguishability between the elements of the corresponding POVM
is 
\(
\Tr \Eop_{1}^{\prime\,\dagger}\Eop_{2}^{\prime}
=
p(1-p)+q(1-q)
\), 
where $\Eop_{n}^{\prime}=\Mop_{n}^{\prime\,\dagger}\Mop_{n}^{\prime}$. 
We stress that all of the features in the measurement operators are
tunable via the basic linear optical elements. 
We also remark that the action of Eqs.\,(\ref{eq:def_gwm1}) and
(\ref{eq:def_gwm2}) can be obtained by a polarizing beam splitter with
tunable reflection coefficients, which has been used for
implementing general two-outcome measurements in optical
setups\,\cite{Huttner;Gisin:1996,Gillett;White:2010}. 

We obtain an important special case of general two-outcome measurements
when either $p=1$ or $q=1$. 
Let us consider the case $q=1$, for example. 
This situation is realized when 
\(
\varepsilon = 1 - 2(r^{\prime})^{2}
\). 
We now find that 
\(
\Mop_{1}^{\prime} 
= 
\sqrt{p}\proj{m_{+}}{m_{+}}
+
\proj{m_{-}}{m_{-}}
\) 
and  
\(
\Mop_{2}^{\prime}
=
\sqrt{1-p}\proj{m_{+}}{m_{+}}
\). 
This is nothing but the partial-collapse measurement of
Refs.\,\cite{Koashi;Ueda:1999,Katz;Korotkov:2008}. 
One possible application of this type of measurements is the proposal by
Korotkov and Keane \cite{Korotkov;Keane:2010} for removing the effects
of decoherence. 

\subsection{General multi-outcome measurements}
\label{subsec:gwmmo}
Next, we examine another generalization of the SASTOM on
$\mathbb{C}^{2}$. 
The repeated application of general two-outcome measurements allows the
construction of a POVM with multiple outcomes. 
Let us consider a system composed of $(N-1)$ detectors, 
each of which performs a general two-outcome measurement. 
Figure \ref{fig:diagram2}(b) shows the case $N=4$. 
At the $\ell$th detector, the first output mode corresponds to
an outcome, while the second output mode is regarded as an
input mode for the subsequent device. 
Thus, we find that the entire system has $N$ outcomes. 
This ``branch structure'' is one possible realization of a multi-outcome
POVM. 
Different geometric arrangements of detectors and paths from
Fig.\,\ref{fig:diagram2}(b) can lead to the same result, as seen in, e.g.,
Ref.\,\cite{Andersson;Oi:2008}. 
In this branch structure, the number of measurements performed changes
from run to run, with an average number $N/2$. 
In the binary-tree structure\,\cite{Andersson;Oi:2008}, the number of
measurements performed is $\log_{2}N$. 

Now, let us show how a general multi-outcome measurement is
implemented. 
Let us write the measurement operators in the $\ell$th apparatus as
$\Mop_{1}^{(\ell)}$ and $\Mop_{2}^{(\ell)}$. 
Their expressions are given in Eqs.\,(\ref{eq:def_gwm1}) and
(\ref{eq:def_gwm2}). 
The measurement operator corresponding to the $\ell$th outcome is
written as $\Kop_{\ell}$ ($\ell=1,\,2,\ldots,N$) and is constructed
recursively using 
\begin{eqnarray}
&&
\Kop_{\ell} 
= \Wop_{\ell}\Mop_{1}^{(\ell)}\Yop_{\ell}
\quad
(2\le \ell \le N-1), \label{eq:def_Nout}\\ 
&&
\Kop_{1} = \Mop_{1}^{(1)}, 
\quad
\Kop_{N} = \Wop_{N}\Yop_{N}.
\label{eq:def_Nout1}
\end{eqnarray}
The linear operator $\Yop_{\ell}$ is defined as 
\(
\Yop_{\ell} = \Mop_{2}^{(\ell-1)}\Yop_{\ell-1}
\) 
($\ell \ge 2$) 
with 
\(
\Yop_{1} = \Iop_{2}
\). 
We remark that $\Yop_{\ell}$ ($2\le \ell \le N-1$) is associated with
the input mode of the $\ell$th apparatus, while $\Yop_{N}$ is
related to the $N$th outcome.  
The unitary operator $\Wop_{\ell}$ is determined by imposing that
$\Kop_{\ell}$ is a positive operator. 
The identity 
\(
\sum_{n=1}^{2}
\Mop_{n}^{(\ell)\,\dagger}
\Mop_{n}^{(\ell)}
=
\Iop_{2}
\) leads to the relation 
\(
 \Kop_{\ell}^{\dagger}\Kop_{\ell}
+
 \Yop_{\ell+1}^{\dagger}\Yop_{\ell+1}
=
 \Yop_{\ell}^{\dagger}\Yop_{\ell}
\). 
This relation indicates the conservation law of probability. 
Using this formula, we show that 
\(
\sum_{\ell=1}^{N}\Kop_{\ell}^{\dagger}\Kop_{\ell} = \Iop_{2}
\). 

A number of interesting quantum protocols with multi-outcome POVM's have
been proposed in the literature. 
Two of the present authors (SA and FN)\,\cite{Ashhab;Nori:2010} proposed
a measurement-only quantum feedback control of a single qubit, for
example. 
Their proposal involves a four-outcome POVM satisfying 
\(
\Tr \Kop_{\ell}^{\dagger}\Kop_{\ell} = 1/2
\) 
and 
\(
\Tr \Eop_{\ell}^{\dagger}\Eop_{\ell^{\prime}}
=
f(x)
\) for $\ell \neq \ell^{\prime}$, 
with $\Eop_{\ell}=\Kop_{\ell}^{\dagger}\Kop_{\ell}$, a continuous real
function $f$, and a real parameter $x$ 
($0 \le x \le 1$). 
We remark that $f$ does not depend on the subscripts 
$\ell$ and $\ell^{\prime}$, but is a function of $x$. 
The real variable $x$ can be understood as the measurement strength. 
An important property of this POVM is that the indistinguishability is
unbiased between arbitrary pairs of elements of the POVM.  
This mutually-unbiased feature can lead to an interesing quantum
control. 
The present procedure is applicable to the construction of the
corresponding measurement operators since one can freely control the
number of outcomes, the trace of $\Kop_{\ell}^{\dagger}\Kop_{\ell}$, and
the indistinguishability. 

\section{Solid-state qubits}
\label{sec:sss}
Let us consider methods for implementing general weak measurements on
solid-state qubits. 
In this paper, we focus on superconducting
qubits\,\cite{You;Nori:2005,Nakahara;Ohmi:2008,Clarke;Wilhelm:2008,You;Franco:2011}. 
Superconducting qubits have many advantages for quantum engineering. 
Their current experimental status\,\cite{Buluta;Nori:2011} indicates
that various important 
quantum operations, especially controlled operations are implemented
reliably. 
The demonstration of controlled-NOT and controlled-phase gates
was reported in various types of superconducting
qubits\,\cite{Buluta;Nori:2011,Yamamoto;Tsai:2003,Plantenberg;Mooij:2007,DiCarlo;Schoelkopf:2009,Neeley;Martinis:2010,deGroot;Mooij:2010,Chow;Steffen:2011}. 
Therefore, it is important for development of measurement-based
quantum protocols to explore the systematic construction methods for
general measurements in such interesting physical systems. 
Several theoretical studies on the implementation of general
measurements in superconducting qubits have been reported in, e.g.,
Refs.\,\cite{Paraoanu:2011:EPL,Paraoanu:2011:FP,Korotkov;Jordan:2006,Ashhab;Nori:2009}. 

Analogies with linear optical qubits are useful for designing
measurement operators in superconducting qubits. 
Let us consider two superconducting qubits, one of which is the measured
system, while the other is an ancillary system. 
The former corresponds to the polarization in the previous arguments,
and the latter is regarded as the path degree of freedom in the
interferometer setup. 
In the interferometer, the polarizing beam splitter plays a central
role to create entanglement between the polarization and the path. 
This operation can be replaced with a controlled operation
(e.g., a controlled-NOT gate) between the two superconducting qubits.  

Now, we show a method for implementing a SASTOM on superconducting
qubits. 
We use the following notation. 
The quantum states of the measured qubit is expressed in terms of
the basis vectors $\ket{+}$ and $\ket{-}$ with 
\(
\ipro{+}{{-}}=0
\). 
The ancillary qubit is described by $\ket{0}$ and $\ket{1}$ with
$\ipro{0}{1}=0$. 
First, we prepare an initial state in the total system 
\begin{equation}
 \ket{\Psi^{({\rm in})}}
=
\ket{\psi}\otimes (\alpha\ket{0}+\beta\ket{1}), 
\label{eq:in_sss}
\end{equation}
with $\alpha^{2}+\beta^{2}=1$, $\alpha,\beta\in\mathbb{R}$, and 
$0 \le \alpha,\beta \le 1$.  
We denote an arbitrary state in the measured qubit as
$\ket{\psi}$. 
The state preparation in the ancillary system can be achieved using 
single-qubit operations. 
Next, we apply the controlled-NOT gate 
\(
\proj{+}{+}\otimes \Iop_{2}
+
\proj{{-}}{{-}}\otimes \hat{\tau}_{x}
\), where 
\(
\hat{\tau}_{x} = \proj{0}{1} + \proj{1}{0}
\). 
The resultant state is 
\(
\ket{\Psi^{(\rm out)}}
=
\Mop_{0}\ket{\psi}\otimes \ket{0} + \Mop_{1}\ket{\psi}\otimes \ket{1}
\), where using $\beta=\sqrt{1-\alpha^{2}}$,  
\begin{eqnarray*}
&&
 \Mop_{0} 
= \alpha \proj{+}{+}
+ \sqrt{1-\alpha^{2}} \proj{{-}}{{-}}, \\
&&
 \Mop_{1} 
=  \sqrt{1-\alpha^{2}} \proj{+}{+}
+ \alpha \proj{{-}}{{-}}. 
\end{eqnarray*}
Therefore, by performing a projective measurement on the state of the
ancillary qubit, we have a SASTOM on $\ket{\psi}$. 
The measurement direction can be changed using single-qubit gates
on the measured system before the controlled-NOT gate.  

The use of a partial controlled-NOT gate leads to the implementation of
general two-outcome measurements. 
Let us now write down the recipe using 
\(
\Uop
=
\proj{+}{+} \otimes \Iop_{2}
+
\proj{{-}}{{-}}
 \otimes \exp(i \xi \hat{\tau}_{x})
\) and the initial state (\ref{eq:in_sss}). 
See, e.g., Ref.\,\cite{deGroot;Mooij:2011} for details of a theoretical
proposal for performing $\Uop$. 
We find that 
\(
\ket{\Psi^{({\rm out})}}
=
\Uop\ket{\Psi^{({\rm in})}}
=
\Xop_{0}\ket{\psi}\otimes \ket{0}
+
\Xop_{1}\ket{\psi}\otimes \ket{1}
\), where 
\(
\Xop_{0}
=
\alpha \proj{+}{+} + (\alpha\cos\xi + i\beta\sin\xi)\proj{-}{-}
\) 
and 
\(
\Xop_{1}
=
\beta \proj{+}{+} + (i\alpha\sin\xi + \beta\cos\xi)\proj{-}{-}
\). 
Depending on the readout result of the ancillary qubit, a proper
single-qubit operation on the measured qubit is performed. 
Then, we find that the state $\ket{\psi}$ is transformed by the positive
operator part of $\Xop_{n}$. 
Using the right-polar decomposition, we obtain the positive operator parts of
$\Xop_{0}$ and $\Xop_{1}$, respectively, 
\begin{eqnarray*}
&&
 \Mop_{0}
=
 \alpha \proj{+}{+} 
+ 
\sqrt{1 - \alpha^{\prime\,2}}
\proj{-}{-}, \\
&&
 \Mop_{1}
=
 \sqrt{1-\alpha^{2}}
\proj{+}{+}
+
\alpha^{\prime}\proj{-}{-},
\end{eqnarray*}
where 
\(
\alpha^{\prime}
=
\sqrt{[1 - (2\alpha^{2}-1)\cos(2\xi)]/2}
\). 

General measurements with multiple outcomes can be implemented in a
similar manner to that given in Sec.\,\ref{subsec:gwmmo}. 
If we obtain the result $0$ in the ancillary qubit, we do nothing. 
A measurement operator [i.e., $\Kop_{1}$ in
Eq.\,(\ref{eq:def_Nout1})] is applied to $\ket{\psi}$. 
Otherwise we perform a single-qubit operation on the measured 
qubit to change the measurement direction and prepare a new superposition
state in the ancillary qubit. 
Then, we apply a partial controlled-NOT gate to the two qubits again. 
Depending on the readout results of the ancillary qubit, we either
obtain one element in the desired POVM [i.e., $\Kop_{2}$ in
Eq.\,(\ref{eq:def_Nout})] or continue to the next step.  
Repeating this procedure, we can obtain any POVM with multiple
outcomes. 
Compared to linear optical qubits, the implementation of a general
multi-outcome measurement in superconducting qubits has an advantage with
respect to scalability. 
In linear optical qubits, it is necessary for the implementation of a
general multi-outcome measurement to prepare all the optical elements
corresponding to all the possible outcomes before the measurement. 
When the number of the outcomes is large, the setup become large and
complicated. 
In addition, most of the elements in the measurement apparatus are
irrelevant to the state in any single run. 
For example, if one obtains the outcome corresponding to
$\Kop_{1}^{\dagger}\Kop_{1}$, the remaining parts of the measurement
apparatus are not used. 
In superconducting qubits, the ancillary qubit can be used in the
different steps of the measurement process. 
In contrast to linear optical setups, the total system is a two-qubit
system even if the number of outcomes is large. 

\section{Summary}
\label{sec:summary}
We have proposed methods for implementing general measurements on a
single qubit in linear optical and solid-state qubits.  
We focused on three types of general measurements on $\mathbb{C}^{2}$. 
The first type is the SASTOM described by Eqs.\,(\ref{eq:def_elm1}) and
(\ref{eq:def_elm2}).    
Their associated POVM is regarded as a minimal extension of a
projection-valued measure. 
The second one is the general two-outcome measurements described by
Eqs.\,(\ref{eq:def_gwm1}) and (\ref{eq:def_gwm2}). 
This is the most general form of the measurements with two outcomes
on $\mathbb{C}^{2}$. 
These two kinds of measurements have only two outcomes. 
Finally, we found that the recursive construction given in
Eq.\,(\ref{eq:def_Nout}) with general two-outcome measurements allows
the design of general $N$-outcome measurements.  

The studies on measurement in quantum mechanics provide 
an interesting research field for both fundamental physics and
applications. 
Systematic and simple methods for the design of general measurements
contributes to the development of this research area. 

\begin{acknowledgments}
We thank P. D. Nation, J. R. Johansson, and N. Lambert for their
 useful comments. 
YO is partially supported by the Special Postdoctoral Researchers Program,
RIKEN. 
SA and FN acknowledge partial support from LPS, NSA, ARO, NSF
grant No.~0726909, JSPS-RFBR contract number 09-02-92114, Grant-in-Aid
for Scientific Research (S), MEXT Kakenhi on Quantum Cybernetics and
JSPS via its FiRST program. 
\end{acknowledgments}

\end{document}